\documentclass[11pt,a4paper,twoside]{article}

\usepackage{amssymb}
\usepackage{latexsym}
\usepackage{multirow,booktabs}
\usepackage{threeparttable}
\usepackage{graphicx}
\usepackage{psfrag,graphicx}
\usepackage{pstricks,pst-text}
\usepackage{amsmath}
\usepackage{natbib}
\usepackage{rotating,booktabs}
\usepackage{tkz-graph}
\usepackage{pstricks,pst-node,pst-tree}%

\usepackage{fancyhdr}

\def\mybibliography#1{{\begin{center} \bf References \end{center}}\list
 {}{\setlength{\leftmargin}{1em}\setlength{\labelsep}{0pt}
\itemindent=-\leftmargin}
 \def\newblock{\hskip .02em plus .20em minus -.07em}
 \sloppy\clubpenalty4000\widowpenalty4000
 \sfcode`\.=1000\relax}
\newbox\TempBox \newbox\TempBoxA
\def\uw#1{%
  \ifmmode\setbox\TempBox=\hbox{$#1$}\else\setbox\TempBox=\hbox{#1}\fi%
  \setbox\TempBoxA=\hbox to \wd\TempBox{\hss\char'176\hss}%
  \rlap{\copy\TempBox}\smash{\lower9pt\hbox{\copy\TempBoxA}}%
}
\newbox\TempBox \newbox\TempBoxA
\def\uwd#1{%
  \ifmmode\setbox\TempBox=\hbox{$#1$}\else\setbox\TempBox=\hbox{#1}\fi%
  \setbox\TempBoxA=\hbox to \wd\TempBox{\hss\char'176\hss}%
  \rlap{\copy\TempBox}\smash{\lower10pt\hbox{\copy\TempBoxA}}%
}
\def\mathunderaccent#1{\let\theaccent#1\mathpalette\putaccentunder}
\def\putaccentunder#1#2{\oalign{$#1#2$\crcr\hidewidth
\vbox to.2ex{\hbox{$#1\theaccent{}$}\vss}\hidewidth}}

\newcommand{\bee}{\begin{eqnarray*}}
\newcommand{\eee}{\end{eqnarray*}}
\newcommand{\be}{\begin{eqnarray}}
\newcommand{\ee}{\end{eqnarray}  }

\def\ni{\noindent}
\def\v{\overrightarrow}

\def\be{\begin{eqnarray*}}
\def\ee{\end{eqnarray*}}

\begin{document}

\thispagestyle{empty}

{\hbox{\footnotesize\rm Journal of Data Science {\footnotesize\bf
 Vol.}}\hfill}

\vspace{0.4pc}

\begin{center}
{\large\bf A Method for Evaluating Options for Motif Detection in Electricity Meter Data }
\end{center}

\vspace{.2cm}
\begin{center}
\renewcommand{\thefootnote}{\fnsymbol{footnote}}

Ian\ Dent$^{1}$, Tony\ Craig$^{2}$, Uwe\ Aickelin$^{1}$, Tom\ Rodden$^{1}$\\~\\

{\it} $^{1}$School\ of\ Computer\ Science\ Nottingham\ University\\$^{2}$The\ James\ Hutton\ Institute\\

\end{center}

{\small
\begin{quotation}
\ni {\it Abstract}:~~Investigation of household electricity usage patterns, and matching the patterns to behaviours, is an important area of research given the centrality of such patterns in addressing the needs of the electricity industry. Additional knowledge of household behaviours will allow more effective targeting of demand side management (DSM) techniques.

This paper addresses the question as to whether a reasonable number of meaningful motifs, that each represent a regular activity within a domestic household, can be identified solely using the household level electricity meter data.

Using UK data collected from several hundred households in Spring 2011 monitored at a frequency of five minutes, a process for finding repeating short patterns (motifs) is defined. Different ways of representing the motifs exist and a qualitative approach is presented that allows for choosing between the options based on the number of regular behaviours detected (neither too few nor too many).

\vspace{0.4cm}
\ni {\it Key words}:~~Motif detection, Clustering, Electricity Usage
\end{quotation}
}

\section{Introduction}
The investigation of household electricity usage patterns is an important area of research given the centrality of such patterns in directly addressing the needs of the electricity industry, both now and in the future. Being able to identify patterns is essential for effective implementation of demand side management (DSM) techniques which itself is necessary to allow the electricity industry to meet the upcoming challenges.

This paper addresses the question as to whether a reasonable number of meaningful motifs, that represent regular activities within a domestic household, can be identified solely using the electricity meter data. The intention is to identify activities that households may do regularly (e.g. the washing of clothes or cooking of the evening meal) and not to identify the usage of particular appliances (e.g. toaster, kettle). The actual physical activity (e.g. washing) will not be determined from the electricity data but the research will determine whether repeating activities can be identified).

The paper adopts an approach of symbolising the electricity meter data and then finding repeating patterns within the symbolised data. Various ways of calculating the symbolised data and various parameter settings are explored to identify the most appropriate parameters for consistently finding repeating activities within the household. The goal is to, firstly, determine whether the data sampled at five minute frequency is sufficient to discover meaningful motifs, and, secondly, to determine the parameter settings that provide the most useful motifs.

The remainder of this paper is structured as follows. Section \ref{background} provides information on the need for changes in analysing electricity usage, how demand side management may address some of the needs, and current work on using motifs to find particular activities. Section \ref{methodology} provides the details on the data used for analysis, the method of finding motifs, and an approach to evaluating whether ``sufficient'' motifs have been found. Section \ref{results} provides the analysis results and Section \ref{conclusions} draws conclusions from the analysis and suggests future work.

\section{Background}
\label{background}

The electricity market in the UK is under various pressures. Some are due to the history and current design of the National Grid and others are arising from worldwide trends, such as the need to reduce carbon emissions and the declining sources of hydro-carbon fuels \cite{Rhodes2010}. New technologies, such as electric cars needing household charging facilities, will be much more common \cite{Poyry2010}. The information monitored in the home will grow rapidly, particularly with the roll-out of Smart Meters planned for UK completion by 2020 \cite{DECC2011}. In addition, the drive to change the mix of electricity generation technologies to reduce greenhouse gas emissions, the desire to reduce carbon dioxide by changing non-electric demand such as gas central heating to the electricity network, and the impact of climate change on altering electricity demand and the greater occurrence of extreme weather events will increase the difficulties in providing a stable and cost effective supply without better modelling of the patterns of consumption within the grid.

\subsection{Demand Side Management}

The electricity market in the UK is undergoing dramatic changes. Legal, social and political drivers require a transformation of existing practices. In particular, the change of sampling of electricity usage from a 3 monthly billing cycle to a 30 minute sampling period using smart meters, alters the degree of understanding of households' behaviour that is possible \cite{Energy2009}.

Historically, electricity supply in the UK has been driven by a desire to provide sufficient supply to match the predicted demand, and to avoid shortages and blackouts. The restrictions on supply due to cost, changing political opinions regarding generation technologies, and international obligations to meet carbon reduction targets, means that, in the future, the emphasis will need to change to demand more closely matching the available supply.  One approach to addressing this issue is the application of demand side management (DSM) techniques to achieve changes in consumer behaviour. \cite{river2005primer} defines DSM as ``systematic utility and government activities designed to change the amount and/or timing of the customer's use of electricity'' for the collective benefit of society, the utility company, and its customers. As domestic usage represents 30\% of the total electricity usage \cite{IainMacLeay2013}, understanding and influencing domestic usage will be a significant part of the solution.

Research in the area of demand side responses has been ongoing for a number of years. For example, \cite{newborough1999demand}, in 1999, demonstrated the ability to reduce UK household peak usage of electricity by up to 60\% by an assortment of interventions including the replacement of some appliances by gas powered equivalents. \cite{Chamberlin1993} also showed that demand side management was being seriously investigated over 20 years ago.

The UK electricity market will allow for more targeted and more complicated tariff offers for customers to provide many benefits including maximising the efficiency of the supply process. \cite{Energy2009} shows that the provision of Smart Meters will allow greatly increased analysis of a customer's electricity usage and provide the ability to make customised offers on pricing and availability to change customer behaviour (for example, to minimise usage during peak periods) or to increase efficiencies in the electricity supply chain in meeting the predicted demand \cite{ofgem2010}.

\subsection{Motifs}

The ability to use the meter data (as collected by smart meters) to discern particular activities can be useful to the electricity supplier in implementing demand side management techniques. If a particular pattern can be interpreted as a particular activity such as cooking the evening meal, then appropriate incentives or penalties can be offered/imposed on the particular household to gain a change in behaviour.

A motif is defined as a previously unknown, frequently occurring pattern within a stream of data. It is distinguished from known patterns which can be found within a data set by a number of well researched methods \cite{Lin2002}. In many cases the repeating patterns are not exactly the same and the power of a motif finding technique is in its ability to find similar patterns that, in real life, represent the same thing. For example, when considering electricity meter data, similar patterns may be discernible that represent the electricity usage during cooking. 

The electricity meter data stream from a household can be considered as a graph of usage against time and regular activities (e.g. cooking) can be seen as similar shaped usage patterns. Short patterns that repeat within the data are defined as ``motifs'' and detection of these motifs, and their timing, can lead to understanding of behaviour within the household.

The pattern finding problem was initially stated by \cite{Lin2002} who distinguished between the problem of efficiently finding defined patterns within a dataset (which the authors felt was generally solved) and the problem of finding repeating patterns within the data which were not previously known and which the authors named as motifs. 

Previous work from \cite{das1998rule} had addressed a related problem of finding rules that relate patterns in one time series to patterns in another. \cite{hoppner2001discovery} considers the problem of finding temporal relationships between primitive patterns in time series in a generalised way. The term ``primitive patterns'' can be defined as motifs.

A significant amount of work has been done in the area of DNA pattern detection and within textual analysis which can be applied to the motif finding problem. If real valued data (as from electricity meters) can be represented by a series of discrete characters (letters) then the techniques of DNA and textual analysis can be applied \cite{castrotime}.

\cite{meo2012lode} is an example of using the frequency of patterns within a sample to derive classification rules. The regular patterns are used to generate a probabilistic model which provides a probability that a particular pattern will occur within a random example of the given class. This approach allows for all patterns to be used at the same time in the classification.

\cite{zeifman2011nonintrusive} reviews the current state of Non Intrusive Appliance Load Modelling (NIALM) which is the technology to identify individual appliance usage from the overall electricity usage. The conclusion from this review is that, currently, no approaches are suitable for detecting all kinds of appliances. The focus in this paper is on finding interesting, repeating patterns of behaviour (e.g. cooking or washing) rather than identifying the usage of individual appliances.

\cite{chang2010load} is a typical study investigating the detection of appliances making use of the overall power usage only - without the need for intrusive appliance level monitoring. This study focuses on the signature of the turning on of the appliance rather than on the total appliance usage load. Like a lot of NIALM studies, some success has been demonstrated in detecting particular appliances but often only in a laboratory environment where the researchers know the actual appliances they are searching for from a short list and/or by making use of intensive monitoring. This monitoring may be at a very high frequency (sub-second) or by measuring multiple items (such as reactive power). This detailed information will not be available from smart meters rolled out across the UK and, while interesting, are seen as of little benefit when considering the wide population.  

\cite{eagle2006reality} has introduced the concept of ``reality mining'' which is defined as the sensing of complex social systems and using various monitoring tools to detect social patterns within routines that are detected. These apply at various time-scales including daily (e.g. getting up, eating lunch), weekly (Saturday sports) and annual (Christmas holiday family visits). While \cite{eagle2006reality} makes use of mobile phones as sensors to detect the routine activities, the concept equally applies to making use of the household meter data readings to detect routines within the household. The paper discusses the amount of entropy in people's lives where ``people who live disorganised lives tend to be more variable and harder to predict''. 

If a particular pattern is seen to recur then the characteristics of the recurrence can be used to drive demand management interventions. For example, if it is detected that a particular household undertakes the same activity at very different times of day, it can be assumed that there may be little or no requirement within that house to do the activity at a particular time and the household may be open to an incentive to change to a time that is more efficient for the overall network. However, a household that shows very regular behaviours (e.g. cooking the evening meal at the same time every day) may be less open to the incentive and the information can be used to help the utility company best target their interventions.

\begin{figure}[ht]
\centering
\includegraphics[width=0.8\columnwidth]{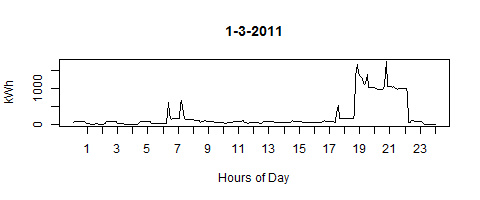}
\caption{Example day of meter readings (household 5 on 1-Mar-2011)}
\label{fig:example-1}
\end{figure}

\begin{figure}[ht]
\centering
\includegraphics[width=0.8\columnwidth]{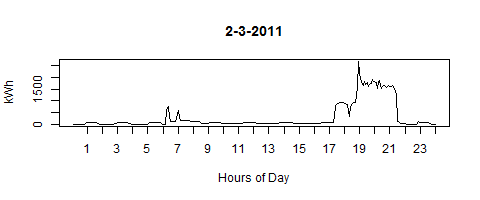}
\caption{Example day of meter readings (household 5 on 2-Mar-2011)}
\label{fig:example-2}
\end{figure}

As an example of what can be interpreted from the shapes of the electricity usage, Figures \ref{fig:example-1} and \ref{fig:example-2} show two days of meter readings for a single house from March 2011. 

Certain aspects of the household behaviour can be assumed as likely from inspection of the graphs:
\begin{itemize}
\item The household members seem to rise from bed at the same time each day
\item It is likely that the household members are out of the house during the working day returning at about 5:00 to 5:30pm
\item The members seem to retire to bed at about 10pm each day
\item There is an underlying repeating pattern throughout the day and night which is likely to be a ``always-on'' device such as a fridge or freezer
\end{itemize}
Note that these assumptions are made from examining only two days of readings but are included to give an example of the types of conclusions that the detection of motifs can help people implementing demand response programs to reach.

The early morning (between 6am and 8am) pattern is repeated approximately on each day and may be a good motif that a larger scale analysis could use to detect the activity of ``rising from bed / breakfast''. Once the motif can be detected on a significant number of days, the variation in timing of the motif can be used to investigate variability of behaviour.

The challenge is for the motifs that the human eye can see as ``similar'' to be detected automatically by a suitable process. The early morning motif can be seen to be slightly different on each of the example days but an observer would use their common sense and knowledge of the types of things people may be doing in the morning (washing, making toast, making tea/coffee) to assume that the patterns show the same type of behaviour. The challenge to be addressed is to find a motif detection method that can automatically match the ``common sense observer''.

\begin{figure}[ht]
\centering
\includegraphics[width=0.8\columnwidth]{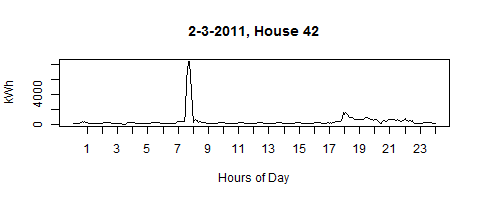}
\caption{Example of large usage for a short period - probably a powered shower}
\label{fig:example-shower}
\end{figure}

Some houses have very high usage for short periods which can distort the shape of the whole day usage (e.g. see Figure \ref{fig:example-shower} which suggests the use of a powered shower). This issue can be addressed by splitting the motifs found within a household into a number of bands (e.g. low range, medium range, high range). If the definition of low and high ranges is taken from the range shown by the household (i.e., 8kW in this case) then only the shower motif will be classed as high range with any other motifs likely to be classed as low range. Instead of taking the definition of ranges from the household minimum and maximum usage, information on typical appliance energy requirements can be used to define low, medium and high ranges and this will allow some of the other motifs within the day (currently dwarfed by the shower) to be identified.

Other work has considered the creation of derived datasets based on the presence (or absence) of particular patterns within the data. For example, \cite{gay2012application} propose a method suitable for classification of data with binary valued attributes. Their work differs from the method proposed in this paper which deals with real valued meter readings and includes a symbolisation step to handle approximate matching of patterns.

The detection of motifs within time series data has been applied to many different application domains but there has been little application of motif detection to electricity meter data from domestic households. What work that has been done has concentrated on activities in the electricity network at a higher level than that of domestic households (e.g. at the local substation level).

\section{Methodology}
\label{methodology}

\subsection{Data selection}

The ongoing North East Scotland energy monitoring project (NESEMP) is examining the relationship between different types of energy feedback and psycho-social measures including individual environmental attitudes, household characteristics, and everyday behaviours.  As part of this ongoing project, several hundred households are being monitored with the electricity usage recorded every five minutes using CurrentCost monitors \cite{Craig2014north}. 

After removing data for households with insufficient readings, the data is loaded into a MySQL database and the readings are aligned with exact 5 minute boundaries (e.g., 1pm, 1.05pm, etc.) by interpolation between the actual readings. A reading at an exact 5 minute point (e.g. 1.05pm) is defined by considering the actual readings before and after that time and by calculating the interpolated reading such that the total usage over a longer period (e.g., an hour) is the same whether the interpolated readings or the original actual readings are used \cite{dent2012}. This results in a set of 288 readings (one for every 5 minute period in the day) for each day and for each of the households in the database. Each day of sampling is labelled in a number of ways such as ``working day'' or ``summer''.

The analysis uses data collected at 5 minute intervals for 123 households from the Spring 2011 period filtered to include just working days (weekdays which are not public holidays). This results in 1,703,232 readings providing a total of 5914 days of readings across all households (not all households have the same number of days of data). 

\subsection{Approach}

Different parameter settings for finding the motifs are used and the results compared using a graphical approach that judges between sets of results (arising from modifying one parameter) and selects the results that would be most useful for a electricity industry professional intending to implement a DSM intervention based around the knowledge of the motif. For example, finding a motif that represents an activity that occurs once or twice a day within a household and then implementing a DSM programme to modify behaviour relating to the activity (e.g. changing the time of occurrence) would be more useful than finding a motif that recurs many time a day (e.g. that arising from a fridge) and which cannot be easily addressed by DSM techniques.  

\subsection{Motif Finding Method}

Various motif detection algorithms are available and this work uses the SAX (Symbolic Aggregate approXimation). Other motif finding algorithms could also be incorporated into the proposed approach (e.g. \cite{mueen2009exact}).

The SAX (Symbolic Aggregate approXimation) technique provides for symbolic representation of time series data and thus provides access to bioinformatics and text mining techniques \cite{lin2007experiencing}. The original work has been extended to iSAX \cite{shieh2008sax} which provides support for large volumes of data. One of the techniques available through SAX is the detection of motifs in a data stream.  

No applications of the SAX technique to domestic electricity data have been found in analysis of the literature although work on wholesale energy price time series has been published by \cite{mori2009sax}.

\begin{figure}[ht]
\vskip 0.2in
\begin{center}
\centerline{\includegraphics[width=\columnwidth]{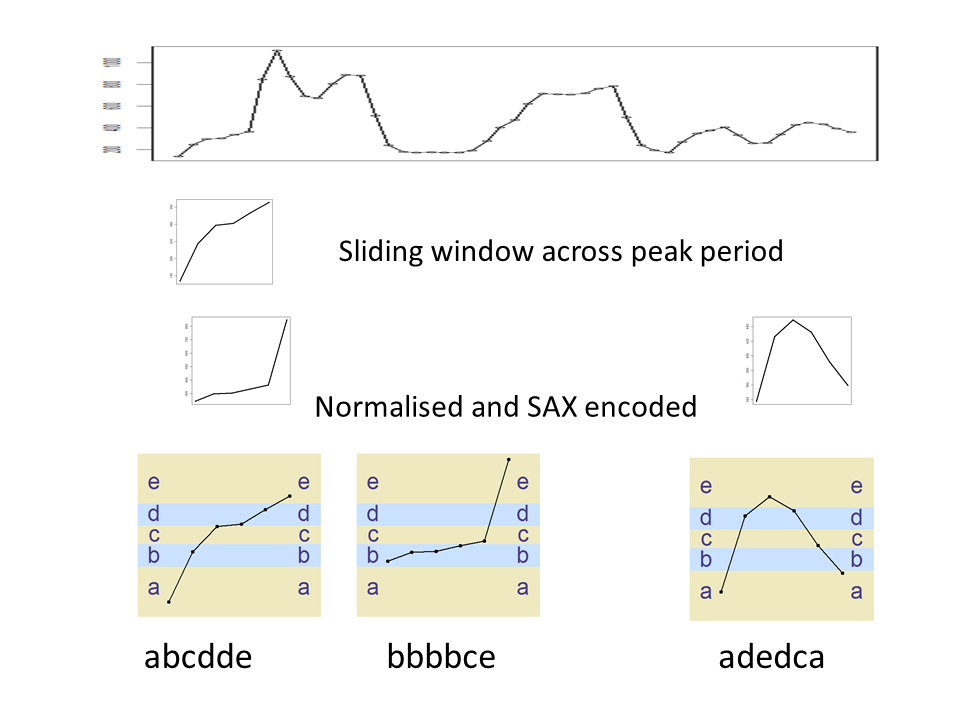}}
\caption{Process for creating SAX representative strings}
\label{sax-explanation}
\end{center}
\vskip -0.2in
\end{figure}

To find the motifs within the data, each period of interest within the day for each household is examined by taking a moving window over the period. The subset of the meter readings within the moving window is then converted into a string using the SAX method and the resulting string is stored in a MySQL database. Next, the window moves on by one reading and the conversion into a SAX representation string is repeated. Thus, using a SAX alphabet size of six and analysing the 4pm to 8pm period (a total of 48 x five minute readings) will result in 43 motifs stored for each day for each household. See Figure \ref{sax-explanation} for an example of how the SAX representations are defined. The top graph shows the five minute readings for a four hour peak period. A sliding window of six readings (30 minutes) is taken across the peak period with the first two windows and the last one shown. Each window is normalised within the values in the window and then translated into the SAX alphabet representations (using an alphabet size of five) as shown at the bottom of the diagram. The example shows a 4 hour peak period but the approach applies to any size for the period of analysis. The analysis in this paper uses the full day as the period of interest.

The initial step in the motif finding process is to symbolise the data. This translates the real valued numbers obtained from the electricity meter into a restricted number of symbolic representations (generally represented as letters). This translation into a set number of symbols (e.g. the letters ``a'' to ``e'' for an alphabet size of five) necessarily incorporates some degree of approximation. As the alphabet size is restricted, the effect is to translate readings within a particular range into a given SAX letter and thus similar, although not identical, readings are translated into the same SAX letter. The resulting SAX string for two motifs may be identical whereas the original readings may only be approximately similar.

Alphabet sizes of 5, 7 and 9 are selected for analysis. An odd number is selected so that no change in meter readings between time points (or only slight changes due to background noise) will map to the ``middle'' character in the alphabet. An even alphabet size would provide two alternatives for the centre symbol and the selection will be driven by noise which is not the desired behaviour.

Too large an alphabet size will lead to very few repeating motifs being found whereas, similarly, too small an alphabet will allow very different activities to be mapped to the same symbolised representation. The selected sizes (between 5 and 9) are analysed further to ensure that a ``reasonable'' number of motifs can be found.

The base motif size used for analysis is 6 which corresponds to a 30 minute (i.e. 6 x five minutes) period. This figure was selected as the UK electricity settlement market uses a 30 minute period and 30 minutes is also a reasonable estimate for a time period that will include most household activities such as breakfast, showering, etc.. 

To explore the setting for motif size that results in a reasonable number of motifs being found, additional motif sizes of 4 (20 minutes), 9 (45 minutes) and 12 (1 hour) are considered to find the most effective setting. Settings for the alphabet size and motif size are considered together to find a reasonable number of motifs.

The motifs found are compared on the basis of similar shape without regard to absolute value of the data.  A possible effect of this is to find motifs within what is the general noise associated with the meter readings. This is avoided by ignoring any motifs within a window which has a range of readings of less than a given size (100W for this analysis).

Various different motifs are defined based on differing ways of considering the data and on how the data is normalised and converted to a symbolic representation. The different motifs created are:
\begin{itemize}
\item That created from considering the actual meter readings, normalised within the moving window being considered, and then converted to a symbolic representation. This process is graphically shown in Figure \ref{sax-explanation}.
\item The difference motif is created by taking the same data from the moving window and then calculating the differences between adjacent meter readings. This provides a series showing the relative increase or decrease in usage from the previous time point. This data is then treated as above (normalisation and symbolisation) to produce the difference motif. 
\item The overall normalised motif is created by considering all the data for each household and then normalising the data within the overall maximum range. The moving window is then taken across this normalised data and each window is then symbolised. Note that the normalisation within the window step is not executed.
\item Compressed motifs are also created from each of the above approaches by removing any repeating letters within the symbolised motif. For example, a motif of ``abcccb'' will be stored in compressed form as ``abcb''.
\end{itemize}

Values for the overall range of the meter readings within the motif window, and the range of the incremental change data (the differences between adjacent meter readings), is stored alongside the motifs.

The final result is that tables specific for each setting of the alphabet size and motif size are created providing for retrieval of the differing types of motifs together with details on the time and date of the motif, the household being considered, and the range information as described above.

The symbolisation step within the motif discovery process is important in allowing for approximate matching. As real valued meter readings are mapped to a few characters, as defined by the alphabet size used, various ranges of value are mapped to the same character and this process has the effect of compensating for some of the noise that is present in meter data collected from households. For example, doing the same activity on different days (e.g. making toast) would be expected to map to the same motif. However, other activities happening at the same time in the house (e.g. the fridge automatically running) or slight differences in the toaster settings (e.g. different level of brownness selected) may lead to slightly different meter readings. With the right symbolisation settings, it is more likely that the activity will map to the same symbolised representation.

It is necessary to detect the repeating motifs that signify particular activities (e.g. cooking the evening meal). These are generally of a similar shape on different days but will show some differences due to noise.

One issue arising is that normalisation means that the shape of the motif is matched without regard to the absolute values of the readings and this might not be valid. While small ranges of less than 100W are excluded, the actual size of the increase or decrease is arguably more useful when identifying particular activities (based on combinations of appliances) rather than the base shape. This can be addressed by normalising over the whole day of a household's readings rather then within the motif window and is the basis for including the normalised difference motif within the data collected.

While the motif matching approach based on shapes can, after normalisation within the analysis window, lead to two motifs being seen as identical when, in fact, one has a much larger range of power readings, this can be addressed by only considering motifs as similar when they fall into the same ranges of power usage.

As the analysis is to find activities during the day, these generally correspond to the use of certain individual appliances (or a combination of appliances) and thus the goal is to find motifs that map onto particular appliance combinations. The power usage characteristics of appliances in common use in the house are given by \cite{Kato2009} (although for Japan rather than the UK).

The Centre for Sustainable Energy \cite{SustainableEnergy2014} publishes a list of typical power usage figures for household appliances. This table can be simplified into the following groups:
\begin{itemize}
\item High usage appliances such as power showers, over 5KW
\item Appliances generally used for heating - water or space - 3KW-5KW
\item Laundry and larger cooking appliances 1KW-3KW
\item Smaller appliances, gardening. Multiple lighting within rooms. 300W-1KW
\item TV, lighting and audio equipment - under 300W
\end{itemize}

The ranges to be used can be set by selecting a reasonable number (e.g. five) and then splitting the range between the lowest and highest reading for a given house into the five ranges. Alternatively, range cut off points can be defined based on general industry standard values for power ranges of common appliances.

The suggested cut-off points in the list above can be used to group the motifs found into similar classes of appliance usage to avoid matching similar shaped patterns that actually cover much different ranges of usage.

\subsection{Identifying ``Interesting'' Motifs}
\label{sec:interesting}

Motifs arising from the small changes in the underlying electricity usage (the ``noise'') are excluded by setting a minimum range of 100W for a motif to be considered of interest. Any motif representing a behaviour which influences the electricity usage by less than 100W over the period of the motif is unlikely to be of interest to professionals designing DSM programmes.

As the motifs are created by shifting a moving window over the stream of data, overlapping periods are considered and a long period with no activity, except for one change in meter reading, will lead to a series of motifs that are similar. For example, when using an alphabet of five, a long period of no activity except for a jump of +200W will lead to motifs such as ccccca, ccccac, cccacc, etc. As only one of these is interesting for further analysis, the others are excluded by omitting any motifs that start with two or more ``c''. With differently sized alphabets the same approach is adopted by excluding the motifs that start with multiple ``middle'' letters.

If a motif consists solely of increases in electricity usage then it is considered to be of no interest. The analysis is attempting to find motifs that reflect complete behaviours (e.g. cooking) and a motif showing only increases in usage would only represent the beginning of the activity. A useful motif will consist of both starting (turning appliances on) and finishing (turning appliances off) an activity. Similarly, a motif which consists solely of decreasing meter readings is also designated as uninteresting.

Finally, any motifs that span midnight are excluded. While this can be criticised as a limitation, it is unlikely that electricity professionals will be considering programmes to change household behaviour around midnight and this limitation is unlikely to affect the usefulness of the results.

\subsection{Evaluation of Different Techniques}

The approach to finding motifs detailed above makes use of a number of parameters such as the type of motif to use (compressed, normalised, etc.), the size of alphabet to use, and the length of the motif. To choose between the various parameters a method is needed to select which set is most appropriate. 

A useful behaviour to focus on for DSM would be one that repeats within a household on the majority of days under consideration and which is of a significant size in terms of power used. It should also be one that only occurs once or a few times per day. An example would be cooking which may only be done once during the conventional time period for the particular meal (e.g. 5pm to 8pm for the evening meal). However, with multiple members of the household, activities may repeat on the same day - e.g. when different groups of the household do their cooking separately.

To detect a repeating activity a reasonable number of motifs need to be found. Some houses will have no repeating activity (their behaviour shows no regularity) and thus no useful motifs may be found. Other houses need to be evaluated in such a way as to give a reasonable number of motifs. The question as to ``what is reasonable'' needs to be addressed. Activity should happen on a sizeable proportion of the days under consideration to be seen as a regular activity. 

\begin{figure}[ht]
\centering
\includegraphics[width=0.7\columnwidth]{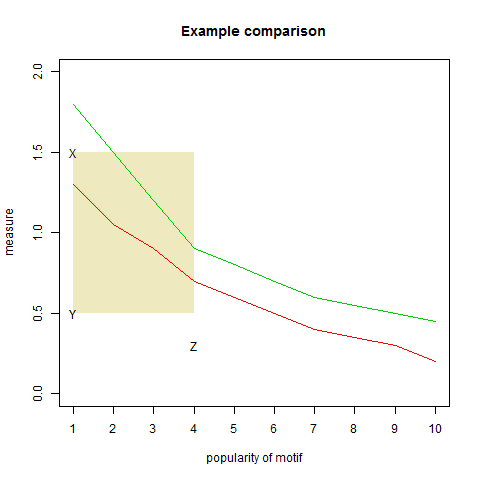}
\caption{Example of motif evaluation: The shaded yellow area is the area of interest - thus the red line is preferred because it falls more into the interesting area than the green line.}
\label{fig:motif_evaluate}
\end{figure}

One approach to evaluation is to consider the x most popular motifs within a household (i.e. the most popular is the one occurring most often, the second most popular is the one occurring second most often, etc.) and then see how the number of occurrences vary between the x motifs. For example, considering the 10 most common motifs, each can be assessed by a useful measure (e.g. number of occurrences) and a graph similar to that of Figure \ref{fig:motif_evaluate} can be produced. The points marked as X, Y, and Z need to be defined in terms of reasonable numbers and then the results from the two sets of parameters displayed on the graph can be assessed against each other.

The point Z sets the number of popular motifs that are to be considered. While the graph shows the results for the top ten motifs, it may be considered that concentrating on the top Z motifs is sufficient and any results to the right of Z can be ignored.

The point X sets the value at which the measure (e.g. number of occurrences per day) is judged to be too high. Finding motifs that occur many times a day (say, ten times) is unlikely to be useful for finding behaviours that occur occasionally and can be influenced. 

The point Y gives a value for the low point of the measure below which motifs are too infrequent to be useful.

These three points define the shaded area which is of ``interest'' and where the useful motif results will lie within. Whichever of the lines on the graphs falls more completely within the shaded area is considered to be the better set of parameters to use for future analysis. In the example, the red line would be judged to give better results as although the green line shows higher values, part of the line falls outside the preferred shaded area and can be considered to give too high results (e.g. too many occurrences per day).

There are many different measures that could be used to distinguish between the motif finding techniques. This analysis makes use of the following:

\begin{itemize}
\item The number of times that the motif occurs within the data period. This should not be too high (point X in the example) nor too low (point Y). 
\item The number of days that include the motif. There is no maximum value (point X) for this measure as a motif falling on every day would be useful. The value for a minimum number of days (point Y) needs to be selected.
\item The percentage of days that include the motif. This differs from the second measure as it incorporates the number of days of readings for the household into the calculation. Not all households have a set of readings for every day within the period of analysis.
\end{itemize}

The relevant values for X, Y and Z need to be set for each of the measures being considered. The values chosen are detailed below. The analysis set out in the remainder of the paper can easily be adjusted for differing values of X, Y and Z and could be explored in future work. For this analysis, values are selected (with justification below) and other values could also be valid.

For each measure, the top three motifs only are considered (i.e. point Z). The value of three is selected as a value between concentrating on too few motifs, and allowing too much influence to motifs that only occur occasionally for a household (e.g. if set as four or more). 

When considering the average occurrences per day for each motif, a value for X needs to be selected that exclude motifs that happen too frequently during the day and which may reflect regular behaviours which the household members may have little opportunity to influence. This could include regular automatic activities (e.g. fridge turning on and off) or behaviours which are necessary for efficient household functioning (e.g. turning lights on and off). A value for X of 2.0 is selected.

In a similar way, motifs that occur infrequently would be of little interest as they do not reflect a regular behaviour and hence Y should not be set too low. A value of Y of 0.3 is selected, reflecting that motifs occurring on average on 30\% of the days is of interest.

Some motifs may occur many times on a few days but rarely on other days and the analysis of the number of unique days that contain the motif is used to identify this situation. The date of occurrence of each motif is found and the number of unique days for that motif is found and then averaged across all the households. In this situation, a motif that occurs on every day under analysis would be of interest and hence X should be set to a maximum value. The value of Y depends on the number of days in the period of analysis and, when considering a calendar quarter of working days (i.e. a maximum number of days of 65), a value for Y of 10 is selected.

Some households do not have readings for every day within the period of analysis and the percentage number of unique days with a given motif provides a measure that takes into account the number of days of analysis. In this case, a figure for X is selected of 90\% which excludes the motifs that occur on nearly every day and may be considered to be too frequent and likely to reflect regular activities which the household has no control over (e.g. fridges). The value for Y is set at 20\%.

The values selected for each measure are summarised in Table \ref{tab:evaluation}.

\begin{table}
\begin{center}
\renewcommand{\arraystretch}{1.3}
\caption{Values selected for each measure}
\label{tab:evaluation}
\begin{tabular}{|c|c|c|c|}
\hline
{\bf Measure} & {\bf X} & {\bf Y} & {\bf Z}\\
\hline
Motifs per day & 2 & 0.3 & 3\\
Total days with motif & 65 & 10 & 3\\
Percent of days with motif & 90 & 20 & 3\\
\hline
\end{tabular}
\end{center}
\end{table}

\section{Results}
\label{results}

The research uses R Studio version 0.97.551 using R 3.0.0 \cite{RCoreTeam2012} running on a Windows 7 64 bit system and accessing the data stored within a MySQL v5.5.31 database. 

\subsection{Exploring the Different Ways of Treating the Motif Range}
\label{sec:range}

Three approaches to handling the range of the motif (the difference between highest and lowest meter readings during the period of the motif) are considered. These are:
\begin{itemize}
\item No consideration of range. Each motif with a similar shape, irrespective of actual values of the meter readings, is considered as the same motif. This approach matches similar shapes but has the possible downside of equating a motif with a relatively small range of readings (e.g. 200W) with the same shape over a relatively large range (e.g. 2KW). It is unlikely that these correspond to the same activity which would generally relate to the use of particular appliances and which would have particular electricity usage characteristics (e.g. Wattage ratings). 
\item Using the range of meter readings within each house under consideration. For each house, the maximum and minimum values for the range of each motif are found. 5 ranges are selected and these are set as being equally spread between the low and high motif range values. For example, for a house with motifs with a maximum range of 1100W and a minimum range of 100W, the ranges are set as 100-300W, 300-500W, 500-700W, 700-900W and 900W to 1100W. Motifs are only considered to be the same if, as well as having the same shape, the range of the motifs match.
\item Based on typical appliance usage characteristics, the range boundaries are set as 300, 1000, 3000, 5000, and 60000W and the motifs are treated as above such that motifs are only judged to match when their shape is similar and they fall within the same range. 
\end{itemize}

\begin{figure}[ht]
\centering
\includegraphics[width=0.9\columnwidth]{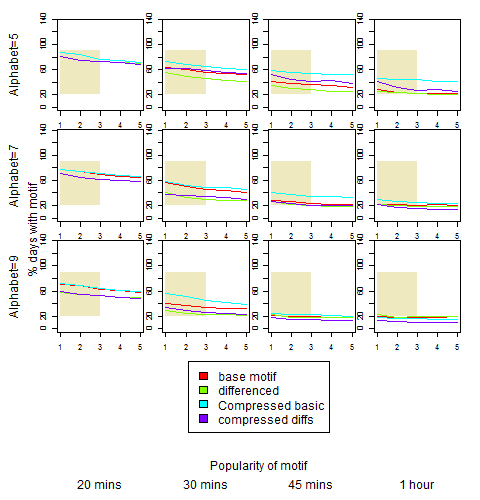}
\caption{Percentage of days; Results with no range adjustment}
\label{fig:motif_range1}
\end{figure}

\begin{figure}[ht]
\centering
\includegraphics[width=0.9\columnwidth]{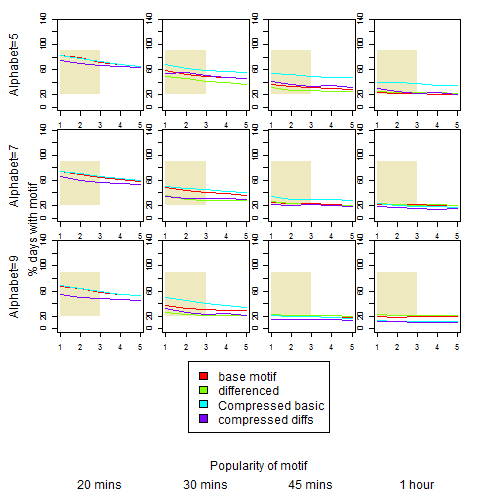}
\caption{Percentage of days; Results with within house range adjustment}
\label{fig:motif_range2}
\end{figure}

\begin{figure}[ht]
\centering
\includegraphics[width=0.9\columnwidth]{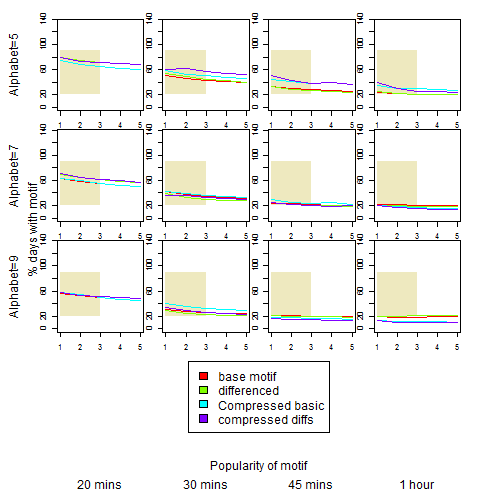}
\caption{Percentage of days; Results with appliance range adjustment}
\label{fig:motif_range3}
\end{figure}

To assess the different approaches to dealing with ranges, the motifs using each approach are found for various values of alphabet size (5,7,9) and motif length (4,6,9,12). The results for each of the motifs detected (using the base data, using differences between readings and the compressed version of each) are plotted with the results shown at Figures \ref{fig:motif_range1}, \ref{fig:motif_range2}, and \ref{fig:motif_range3} for the percentage of days where the motif is found. The other measures (motifs per day and number of unique days) are also considered and the results are similar (graphs not included for space reasons). The graphs show the results for each value of alphabet size (5,7,9) down the vertical and each value of motif size (4,6,9,12) across the horizontal.

Except for the largest values for motif size and alphabet size (where, intuitively, fewer motifs would be expected) all approaches to dealing with the range provide results within the acceptable and useful area (the shaded areas on the graphs). As the approach of making use of the ranges of possible appliances (the third approach) is intuitively more appealing, this option is taken for future analysis. 

\subsection{Evaluating Normalisation Method}

Options considered for normalisation of the data prior to conversion to symbolised representation are:
\begin{itemize}
\item Each motif is considered and the values within the motif are normalised before conversion to a symbolised string.
\item Each household is considered and all the readings for that household are normalised prior to consideration of motifs.
\end{itemize}

\begin{figure}[ht]
\centering
\includegraphics[width=0.9\columnwidth]{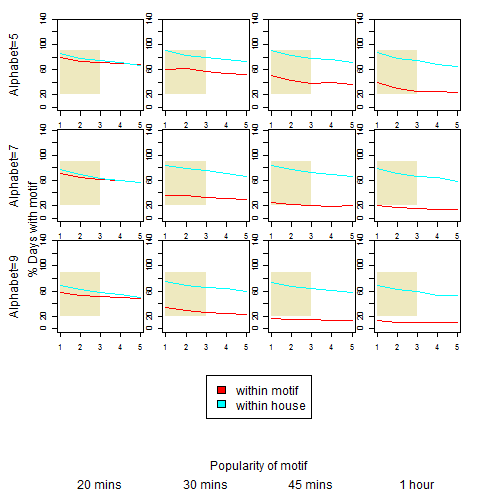}
\caption{Unique days (\%) for different normalisations}
\label{fig:motif_norm1}
\end{figure}

\begin{figure}[ht]
\centering
\includegraphics[width=0.9\columnwidth]{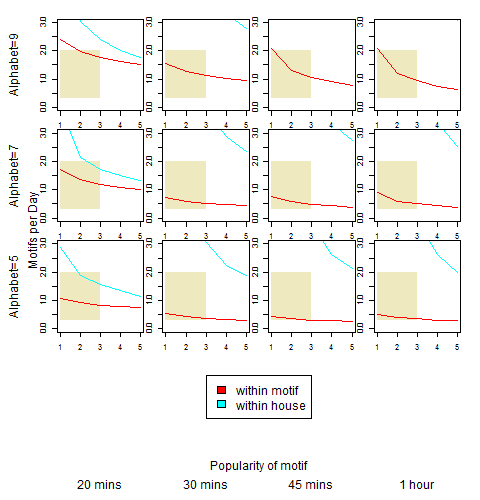}
\caption{Motifs per day using different normalisations}
\label{fig:motif_norm2}
\end{figure}

The results from considering the two approaches to dealing with normalisation are shown in Figures \ref{fig:motif_norm1} and \ref{fig:motif_norm2}. Again the results for the number of unique days is omitted for space reasons.

The motifs per day results show that the normalisation over the whole period for each household produces a lot of motifs per day and, in fact, more than the maximum 2 per day that is considered useful. For that reason, the normalisation within a motif is considered the better approach and is selected for further analysis.

\subsection{Selecting the Motif Size}

\begin{figure}[ht]
\centering
\includegraphics[width=0.9\columnwidth]{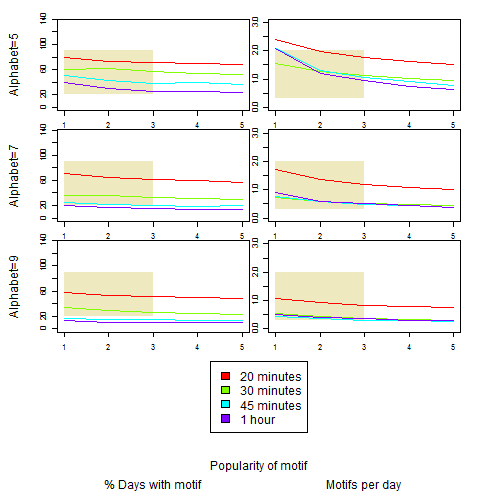}
\caption{Different motif sizes}
\label{fig:motif_size}
\end{figure}

As the motif size increases, the number of possible combinations of letters within the motifs increases and the number of motifs found reduces. The requirement for a useful size of motif is to provide useful numbers of motifs (but not too many). To analyse possible settings, results for motifs of size 4, 6, 9, and 12 (corresponding to 20 minutes, 30 minutes, 45 minutes and an hour) are each used with alphabet sizes of 5, 7 and 9. The results are shown in Figure \ref{fig:motif_size}.

For smaller values of alphabet size, the 30 minute motifs give more useful results. However, for larger alphabet sizes the 20 minute motifs are better. To investigate this further, both the 20 minute and 30 minute motifs will be considered when evaluating the alphabet size.

\subsection{Selecting the Alphabet Size}

\begin{figure}[ht]
\centering
\includegraphics[width=0.9\columnwidth]{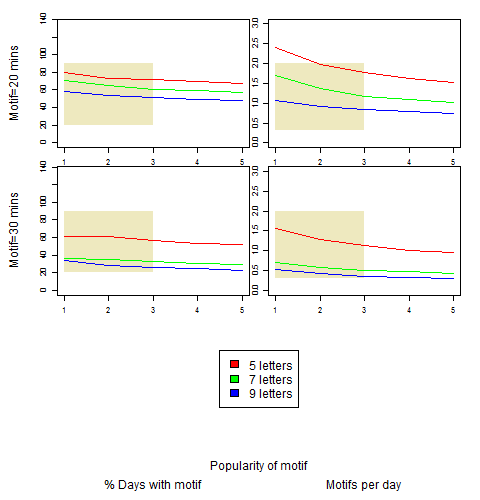}
\caption{Different alphabet sizes}
\label{fig:alphabet_size}
\end{figure}

As the alphabet size increases, the number of possible combinations of letters increases and the number of times motifs repeat will be expected to decrease. To evaluate the best settings for alphabet size, each of 5, 7, and 9 letters are considered with the motif sizes of both 4 and 6 (20 minutes and 30 minutes). The results are shown in Figure \ref{fig:alphabet_size}.

The results show that two combinations of settings are likely to provide useful motifs for further analysis. When using a small number of letters (5) in the alphabet when symbolising motifs, the best motif size to use is 6 equating to 30 minutes. However, an alphabet size of 7 is best used in combination with a motif size of 4 (20 minutes).

\section{Conclusions and Further Work}
\label{conclusions}

Based on the above analysis, a set of parameters is selected as the ones that are most likely to provide motifs that will be useful. The parameter settings are shown in Table \ref{tab:parameters}.

\begin{table}[ht]
\small
\caption{Parameter Settings selected for future analysis}
\begin{center}
\begin{tabular}{|p{0.2\textwidth}|p{0.3\textwidth}|p{0.4\textwidth}|}
\hline\noalign{\smallskip}
\textbf{Parameter} & \textbf{Setting}  & \textbf{Meaning} \\ \hline
Range & Using appliance characteristics & The typical ranges of appliances can be used to distinguish between similar shaped motifs.\\  \hline
Data & Differences between consecutive readings & The differences between readings are changes in electricity usage.\\ \hline
Compression & Compressed data used & Repeating letters within the motif are removed.\\ \hline
Normalisation & Within motif & Values within the motif are normalised before being symbolised. \\ \hline
Motif size & 4 (20 minutes) or 6 (30 minutes) & The length of each motif.\\ \hline
Alphabet size & 5 or 7 letters (depending on motif size) & The number of letters used for the motif symbolisation step.\\
\hline \end{tabular}
\label{tab:parameters}
\end{center}
\end{table}

The combination of parameters has been shown to produce a usable number of motifs. The conclusion is therefore drawn that using a sampling frequency of five minutes is sufficient for repeating patterns to be detected within the data streams and that sufficient meaningful patterns can be found for the motif analysis to be useful.

\begin{figure}[ht]
\centering
\includegraphics[width=0.9\columnwidth]{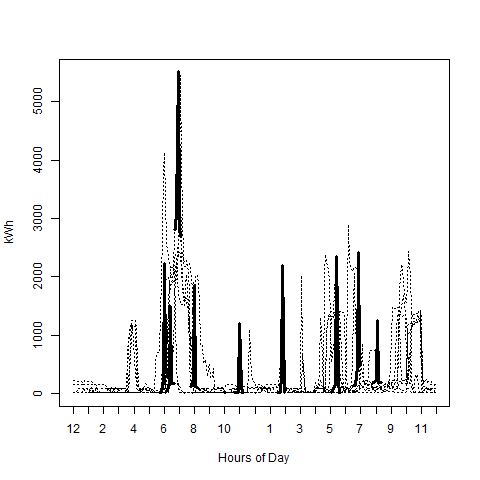}
\caption{Example of motifs found for house 485}
\label{fig:example}
\end{figure}

An example showing the motif found for house 485 (where there is also a large range for the readings within the motif) can be seen at Figure \ref{fig:example}. This shows the daily usage pattern for each day (each dotted line) for the selected house within the Spring 2011 period (note that this house has data for about 350 days out of 365 days of posible readings) with the particular motif picked out in bold. The motif broadly represents a large spike in usage preceded and superseded by a period of relatively little change in activity. It can be seen that the motif is found even though the spike in usage may be a slightly different height on each day (so long as the overall range fits within the 1000-3000W range). It can also be seen that the motif at about 7am is detected even though there is other electrical activity on that day providing a base load onto which the motif usage is added. Other spikes in usage that have not been found to match the motif (i.e. not represented in bold) generally show a gradual tail-off in usage rather than a sharp reduction to the original level of usage and probably represent different household activities.

This paper has addressed the question as to whether a reasonable number of meaningful motifs, that represent regular activities within a domestic household, can be identified solely using the electricity meter data. The analysis has shown that data sampled at a five minute interval is sufficient to provide a usable number of motifs which can be assumed to represent repeating activities within the household. 

The symbolisation process is a suitable method for converting the real valued meter readings into a restricted set of characters with each representing a certain range of values. This provides for matching of data samples which, while not identical, map to the same symbols and can be assumed to be approximately the same.

The results show that motif finding using the symbolisation technique and the five minute sampling period is possible. A series of different parameter settings have been explored and a set of parameters selected.

The method of representing the motif in an alphabetised form has been explored with possible approaches of using the base meter data or differenced data. In addition, compression of repeating characters is considered.

Motifs may match but actually represent similar patterns within very differently sized data and different approaches to splitting the motifs into various bands of usage over the range of the motif have been explored with a solution based on a split into various appliance ranges (five ranges from small to large usage) selected as the best approach.

Various motifs that show uninteresting behaviour can be found (e.g. no change in usage over the time of the motif such as when the house is unoccupied) and an automatic method of rejecting some of the uninteresting motifs has been developed.

An effective method of identifying meaningful motifs opens up opportunities for electricity industry professionals to gain greater understanding of activities within given households. The paper has not tested the implementation of initiatives to change behaviour but has investigated whether reasonable numbers of motifs can be identified and has established the most appropriate parameter settings for obtaining reasonable numbers. Future work can make us of the derived parameters settings to explore the detection of behaviours and then test interventions to change these behaviours. This would be a very valuable step along the road of effectively deploying effective DSM initiatives.

%
%





\bibliographystyle{ECA_jasa}
\bibliography{References}



\end{document}